# Enhanced orbital electron-capture nuclear decay rate in compact medium

A. Ray[1], P. Das[1], S. K. Saha[2], A. Goswami[3], A. De[4]

[1]Variable Energy Cyclotron Centre, 1/AF, Bidhan Nagar, Kolkata – 700064, India

[2]Radiochemistry Division, Variable Energy Cyclotron Centre, 1/AF, Bidhan Nagar, Kolkata – 700064, India

[3]Saha Institute of Nuclear Physics, 1/AF, Bidhan Nagar, Kolkata – 700064, India

[4]Raniganj Girls' College, Raniganj, West Bengal, India

Abstract: The eigenstate energies of an atom increase under spatial confinement and this effect should also increase the electron density of the orbital electrons at the nucleus thus increasing the decay rate of an electron capturing radioactive nucleus. We have observed that the orbital electron capture rates of $^{109}$In and $^{110}$Sn increased by (1.00±0.17)% and (0.48±0.25)% respectively when implanted in the small Au lattice versus large Pb lattice. These results have been understood because of the higher compression experienced by the large radioactive atoms due to the spatial confinement in the smaller Au lattice.

## I. Introduction

It is known from the earlier works that the decay rate of electron-capturing $^7$Be changes a little bit in different environments such as in different beryllium compounds [1-3] or when implanted in different media such as Au, graphite, $Al_2O_3$, $LiO_2$, fullerene ($C_{60}$) etc. [4-9]. A maximum change of slightly over 1% was observed by comparing the decay rates of $^7$Be in all these media. These results have been qualitatively understood [10] in terms of the electron affinity and lattice structure of the host media causing change of the number of valence 2s electrons and hence the electron density at the nucleus. Quantitative estimates [10] have also been obtained using the Tight Binding Linear Muffin-tin Orbital Method (TB-LMTO) [11] and Hartree and Hartree's calculations [12]. Reasonable agreement with the observations has been found.

Since only s-electrons can have significant overlap at the nucleus and this overlap also drops rapidly for higher orbital s-electrons, no significant change of decay rate in different environments was generally expected for higher atomic number electron-capturing nuclei. Norman et al. [5] searched for the change of decay rate of electron capturing $^{40}$K in different potassium compounds and did not find any effect within 1%. Isomeric $^{99}$Tc decays by internal conversion of valence shell electrons and a small change [13] (≈0.3%) of its decay rate was observed in different chemical environments. However for the vast majority of electron-capturing nuclei, valence shell electrons do not have any significant interaction with the nucleus.

The decay rate of an electron-capturing nucleus can also be increased by applying external pressure on the radioactive atom. Hensley et al. [14] applied a high pressure of up to 270 kilobar to $^7$BeO and found that the decay rate of $^7$Be increased by 0.59%. Recently Liu and Huh [15] found that the decay rate of $^7$Be(OH)$_2$ increased by 0.88% when about 270 kilobar pressure was applied on it. As a result of applying external pressure, lattice volume decreases and $^7$Be ion is confined to a smaller spatial volume thus increasing the eigenstate energies of orbital electrons and the electron density at the nucleus. Similarly ions implanted in different lattice environments should also experience different levels of compression affecting the electron density at the nucleus and the corresponding electron capture rate. However the differential compression experienced by $^7$Be in different lattice environments under normal circumstances (i.e. without the application of substantial external pressure) is generally negligible due to the small size of the beryllium atom and so the decay rate of $^7$Be implanted in different lattice environments essentially depends on the available number of valence 2s electrons of $^7$Be, as discussed in our earlier works [4, 10]. On the other hand, the compressional effect of the lattice environment may not be negligible when a much bigger atom such as indium or tin is implanted. So, the electron capture decay rates of $^{109}$In and $^{110}$Sn might increase in smaller lattices such as Au and Al compared to that in a larger lattice such as Pb. Although the lattice sizes of Au and Al are similar, but Al is a smaller atom than Au and so the available octahedral and tetrahedral spaces are smaller in Au lattice than in Al lattice. Hence, the electron-capture decay rates of $^{109}$In and $^{110}$Sn should be faster in Au compared to Al.

**II. Experiment**

We report here our measurements of the half-lives of $^{109}$In ($\tau_{1/2} \approx 4.2$ hours) and $^{110}$Sn ($\tau_{1/2} \approx 4.2$ hours) implanted in gold, aluminum and lead foils. $^{109}$Sn ($\tau_{1/2} \approx 18$ minutes) and $^{110}$Sn ($\tau_{1/2} \approx 4.2$ hours) nuclei were produced by bombarding a 10.7 mg/cm$^2$ thick niobium foil with a 150 MeV $^{20}$Ne beam from the Variable Energy Cyclotron Center, Kolkata, India. The average $^{20}$Ne beam current was about 2.5 pnA and the duration of each irradiation was 8-10 hours. $^{109}$Sn ($\tau_{1/2} \approx 18$ minutes) and $^{110}$Sn ($\tau_{1/2} \approx 4.2$ hours) nuclei thus produced via one proton plus three neutron emissions and one proton plus two neutron emissions came out from the niobium foil with an average kinetic energy of about 20-25 MeV and were implanted in a catcher foil placed behind the niobium target. We used a 50 mg/cm$^2$ thick gold foil, a 23 mg/cm$^2$ thick lead foil and a 6.8 mg/cm$^2$ thick aluminum foil as catchers. The ranges of 20-25 MeV tin or indium ion in gold, lead and aluminum are between 3-4 µm, whereas the range of about 80 MeV $^{20}$Ne that came out from the back of the niobium foil was around 15-25 µm in those media. According to TRIM code calculations [16], the $^{20}$Ne beam used for our irradiation work was expected to produce little damage ($\approx 0.01$ vacancies/ion/Angstrom) in the lattice sites where 20-25 MeV tin or indium ions would stop. The estimated number of damaged lattice sites should not be more than a few percent of the total number of lattice sites and so the change of decay rate in different media should be primarily determined by the properties of undamaged lattice sites.

The amount of energy deposited by beam at the catcher foil was about 0.2 Watt. Considering both the radiation and conduction losses, the corresponding temperature rise of the beam spot region is estimated to be less than 5°C for gold and aluminum foils and less than 50°C for lead foil. Some loss of material from the lead catcher foil due to the beam hitting was also observed. The implanted indium and tin ions were expected to be in equilibrium octahedral and tetrahedral positions of the lattice as the catcher foils reached room temperature immediately after irradiation. There might be slight annealing effect in lead, but the number of damaged sites as a fraction of total number of atoms is also somewhat higher for lead.

After each implantation run, the radioactive ion implanted foil was cooled for an hour so that all short-lived (half-life of the order of seconds and minutes) radioactive nuclei would decay. Then the foil was counted by placing it in front of a four-segmented CLOVER detector manufactured by Canberra-Eurisys. The details about the detector are given in ref [17]. The distance of the source from the front of the detector was about 5 mm. In order to correct for dead time of the measurements and other systematic errors, a $^{60}$Co source ($\tau_{1/2} \approx$ 5 years) was also placed at a fixed distance (about 30 cm) from the detector assembly. $^{110}$Sn decays by electron capture and a 280 keV γ-ray is emitted from the daughter $^{110}$In nucleus. $^{109}$Sn nucleus decays to $^{109}$In by positron emission and electron capture with a half-life of about 18 minutes. Then the $^{109}$In nucleus decays predominantly by electron capture and a 203 keV γ-ray photon is emitted from its daughter $^{109}$Cd. The ratio of the peak area of the 280 keV γ-ray to the sum of the peak areas of 1173 keV and 1332.5 keV γ-rays of $^{60}$Co was monitored with time to determine the half-life of $^{110}$Sn nucleus. Similarly, the ratio of peak area of the 203 keV γ-ray to the sum of the peak areas of 1173 keV and 1332.5 keV γ-rays of $^{60}$Co was monitored with time to determine the half-life of the $^{109}$In nucleus. In order to allow almost complete decay of $^{109}$Sn to $^{109}$In ($\tau_{1/2} \approx 18$ minutes), we waited for about 3 hours after irradiation to start monitoring the 203 keV γ-rays produced due to the electron capture of $^{109}$In nucleus. The time was kept using a precision pulser whose signal was sent to a CAMAC scaler. The single scaler counts and four spectra from the CLOVER detector were acquired for successive intervals of 15 minutes duration and then written on a computer disk. This was followed by an automatic reset of the scalers, the erasure of the spectra from the spectrum buffer and the start of data collection for the next 15 minute interval. The livetime of the counting system increased with time as the short-lived sources cooled down. However the ratio of two peak areas should be independent of the livetime of the counting system and this was verified by monitoring the ratio of the peak areas of 1173 keV to 1332.5 keV γ-ray lines from $^{60}$Co with time.

**III. Analysis and Results**

In Fig. 1, we show typical γ-ray spectra from the radioactive ions implanted in a gold and in a lead foil. We find 203 keV and 280 keV γ-ray lines from the electron captures of $^{109}$In and $^{110}$Sn and also γ-ray lines from $^{60}$Co. All other γ-ray lines having intensity greater than or equal to 0.05% of the most dominant 203 keV peak have been identified. We have studied 80 γ-ray lines in each spectrum spanning the energy region from 60 keV to 1600 keV. Apart from a few background lines, all of them came from the reaction products (mostly $^{109}$In, $^{110}$Sn and $^{110}$In) of $^{20}$Ne on $^{93}$Nb. None of those 80 γ-ray lines can have associated lines that can contaminate our regions of interest. In the case of the aluminum catcher, 270 keV, 372 keV, 1158 keV and 1501 keV lines from $^{44}$Sc, $^{43}$Sc and $^{44}$K were also seen. The only possible contamination can come from the electron capture decay of $^{110}$In, because it has an approximately 1% γ-ray branch [18] at 1334 keV that will interfere with one of the $^{60}$Co peaks used for normalization. Using the intensity of the 642 keV line (having 43.8% branching ratio) [18] from the electron capture decay of $^{110}$In, we estimate about 0.01% overestimation of indium and tin half-lives due to this contamination. No other reaction product or background line can contaminate our regions of interest. Using Gamma-Ray catalog [19], we also considered all known γ-ray lines that can contaminate our regions of interest. The half-lives of those isotopes are either too long or too short and the associated γ-ray lines have not been seen in our spectra. For example, $^{107}$In (half-life = 32.4 minutes) undergoes electron capture decay and produces 205 keV γ-ray line that can contaminate our region of interest. However it should also produce 320.9 keV and 505 keV lines, but none of them have been seen. It may be hard to see 505 keV line due to the presence of very strong and broad 511 keV line, but 320.9 keV line should have been seen easily, if it is produced. Similarly $^{105}$Ag isotope (half-life = 41.3 days) can produce 280 keV γ-ray line that can contaminate our region of interest. Then the associated 443.4 keV γ-ray line should also be seen, but it has not been seen. In this way, we have convinced ourselves that there is no contamination in our regions of interest that can affect our measurement at the level of 0.1%. From Fig. 1, we also find a prominent 658 keV γ-ray line from $^{110}$In decay. It would have been nice to be able to carry out decay rate study of this peak to determine the half-life of $^{110}$In and how its decay rate is affected in different media. However the 658 keV γ-ray line is coming from two sources a) decay of $^{110}$In (ground state) having half-life = 4.9 hours and b) decay of $^{110}$In (2$^+$ state) having half-life= 69.1 minutes. Because of its composite character, this γ-ray line is not suitable for high precision half-life measurement. The initial ratio of the populations of $^{110}$In (ground state) to $^{110}$In (isomeric 2$^+$ state) is unknown and could be different for different irradiation runs due to the small changes in cyclotron beam energy, thickness of niobium foil etc.

The peak areas were determined by subtracting out a linear background drawn under the peak by joining two points on the two sides of the peak and exponential fits of the decay curves were performed. We find that the quality of the exponential fits and the results were essentially independent of the peak area determination method as long as any reasonable method is used consistently throughout the analysis. In Fig. 2, we show the exponential fits for the data points of $^{109}$In implanted in Au and Pb as well as the corresponding residual plots obtained by subtracting out the calculated fitted points from the actual data points. The uncertainties of the data points have been calculated using counting statistics only. The corresponding reduced $\chi^2$ values as shown in Table 1 are somewhat larger than 1.00 indicating that the errors have been underestimated. In order to study the nature of the additional errors, we have done linear fits of the residual data points using the linear function $f(t) = a - bt$, where t is time and a, b are constants. The results of the fits are a= $(2.63 \times 10^{-3} \pm 1.10 \times 10^{-1})$% and b=$(1.70 \times 10^{-5} \pm 2.92 \times 10^{-4})$%min$^{-1}$ for $^{109}$In in Au case and a = $(1.04 \times 10^{-2} \pm 1.35 \times 10^{-1})$% and b = $(4.57 \times 10^{-5} \pm 4.33 \times 10^{-4})$%min$^{-1}$ for $^{109}$In in Pb case. These results are consistent with null results and the reduced $\chi^2$ values of the straight line fits are 1.88 and 1.94 for $^{109}$In in Au and $^{109}$In in Pb residual data point plots (Fig. 2) respectively. We have also done Legendre polynomial ($P_\ell(t)$) fits of the residual data points to study the nature of the apparent wavy structures of the residual data points. For this purpose, the time co-ordinates were suitably normalized so that their values lie between 0 to 1. Then the residual data points have been fitted with Legendre polynomials $P_\ell(t)$ for $\ell$ = 5,10,20,35 and 50. The corresponding values of reduced $\chi^2$ become slightly higher than the corresponding linear fit values in most cases and remain about the same in other cases. So these results indicate that there is no statistically significant structure in the residual plots. We have also done double exponential fits of the form $f(t) = ae^{-bt} + ce^{-dt}$ of the original data points for both $^{109}$In in Au and Pb cases to search for the possibility of decay of any small impurity line that we might have missed during spectral analysis. The double exponential fit does not improve the quality of the fits and in some cases make it slightly worse. We find that the uncertainties on both the amplitude and decay constant of impurity line is much larger than the values of the corresponding amplitude and decay constant. So our detailed analysis is not showing any indication of systematic errors from the residual plots and double exponential fits. We have also studied the ratios of the peak areas of 1173 keV to 1332.5 keV γ-ray lines from $^{60}$Co with time and found that the data points can be best fitted by straight lines parallel to the time axis. However the fluctuations of the data points have been found to be greater than just the counting statistics and give reduced $\chi^2$ values similar to those obtained by fitting residual data points of Fig. 2. From all these results, we conclude that the statistical errors associated with the determination of peak areas have been underestimated, but the additional error is clearly statistical in nature. We have only considered the uncertainties due to the counting statistics for the determination of peak areas, but there are also random errors associated with setting windows around different peaks for extracting their areas. The total statistical error for the determination of a peak area should include both the uncertainties due to the counting statistics as well as the random error associated with setting window around the peak for extracting the area. These additional random errors have been included in our final results by multiplying our extracted uncertainties of half-life values by the square root of the corresponding reduced $\chi^2$ value.

In Fig. 3, we show the superimposition of two exponential fits for $^{109}$In in gold and lead and the divergence of the two exponential fits after a long time. For the purpose of making the superimposition plot, the initial numbers of $^{109}$In nuclei in Au and Pb were determined from the corresponding decay curves, appropriately normalized and the corresponding data sets of the decay curves were multiplied by the normalization constants to obtain the superimposition plot. Taking the half-life of $^{60}$Co = (1925.28±0.14) days [20], we finally obtain the half-lives of $^{109}$In and $^{110}$Sn in gold, lead and aluminum as shown in Table 1. Both the Au and Pb implantation runs were repeated and the half-lives of $^{109}$In and $^{110}$Sn in Au and Pb were re-measured. The results from both the sets agree well within statistical error bars. Although our measurement should not be affected by the dead time of the system, because we monitored the ratio of two peak areas, but still in order to test the effect of the count rate on half-life measurement, we split the data set in two parts — a high count rate data set (having integrated count rate ≈ 8000 cps) and a relatively lower count rate data set (having integrated count rate ≈ 4000 cps). The half-life measurements from the two sets agree well within statistical error, implying no significant systematic error due to different count rates. The final half-life numbers given in Table I, were obtained by taking weighted average of the results obtained from different runs. The reduced $\chi^2$ values shown in Table I, were obtained by considering statistical uncertainties on counting statistics only. Our detailed analysis (as described earlier) does not show presence of any systematic error in our measurements. So we are justified to increase the statistical errors associated with the fitted half-lives by the square root of the reduced $\chi^2$ values to account for the somewhat larger values of reduced $\chi^2$, as discussed earlier.

Our much higher precision values of half-lives (given in Table I) are in agreement with the previous measurements [21] of half-lives of $^{110}$Sn ($\tau_{1/2}$ = 4.173±0.021 hours) and $^{109}$In ($\tau_{1/2}$ = 4.167± 0.018 hours). As reported in Table 1, the half-lives of $^{109}$In and $^{110}$Sn in gold are shorter than those of $^{109}$In and $^{110}$Sn in lead by (1.00±0.17)% and (0.48±0.25)% respectively. Compared to the half-lives of $^{109}$In and $^{110}$Sn in aluminum, the half-lives of $^{109}$In and $^{110}$Sn in Au are shorter by (0.21±0.20)% and (0.25±0.22)% respectively.

**IV. Theoretical Calculations and comparison with data**

The lattice structure [22] of both gold and lead is face centered cubic, but the lattice constant of gold is 4.08 Angstrom, whereas that of lead is 4.95 Angstrom. The valence orbitals of tin and indium are $5s^2 5p^2$ and $5s^2 5p^1$ respectively and the electron affinities [22] of gold and lead are 222.8 kJ/mol and 35.1 kJ/mol respectively. We have done quantitative calculations using the Tight binding linear Muffin-tin orbital method (TB-LMTO) code [11]. In this code, implanted ions were placed at octahedral or tetrahedral positions and density functional calculations were carried out [10,11]. According to our TB-LMTO calculations [10,11], a tin or indium atom should lose more valence 5s electrons (about 0.5 electrons on the average) when implanted in Au versus Pb lattice. However the corresponding change of electronic overlap at the nucleus (considering both direct overlap and change of shielding effect on inner shell electrons) is negligible and cannot account for the observed effect.

However both tin and indium atoms are expected to experience higher level of compression in Au lattice compared to that in Pb lattice. As a result of this compression, electron density at the nucleus would increase and correspondingly the electron-capture rate would also increase. In the cases of tin and indium atoms, all electrons up to principal quantum number n=4 are treated as core electrons by the TB-LMTO code. A frozen core picture implying no interaction between the core and the lattice environment is assumed. TB-LMTO code [11] considers interactions among the valence electrons of neighboring atoms resulting in the hard sphere approximation [11]. In this approximation, the atoms can be treated as hard spheres of potential (known as repulsive Hartree potential) packed together in a lattice. As a result of treating atoms as hard spheres of potential, the eigenstate energies of core electrons increase when the atom is in a lattice environment. From the study of X-ray photoelectron spectroscopy of different compounds and metals, core level electronic binding energy shift was seen earlier [23]. So such effect is expected for implanted ions also. Since the atomic electrons are constrained inside a hard sphere (called muffin-tin sphere), so the atomic orbitals should be more compressed and the eigenstate energies of the orbitals should increase as the muffin-tin radius is reduced. Assuming that the implanted ion would go to tetrahedral and octahedral sites, we find from TB-LMTO calculations that on the average, the eigenstate energies of 1s, 2s, 3s, 4s orbitals of tin increase by 26.9 eV, 27.0 eV, 25.8 eV and 21.3 eV respectively when tin is implanted in Au versus Pb lattice. The calculated values of eigenstate energies depend to some extent on the initial choice of the muffin-tin radius of the implanted ion and can be varied by about 20% - 25% for different initial choices of the muffin-tin radius. The code does not run if the initial choice of the muffin-tin radius is made too small or too large. TB-LMTO calculation also predicts, on the average, an increase of about 0.4 electrons in the n=5 valence shell of a tin ion implanted in Au lattice compared to that implanted in lead lattice. Such increase in the number of n=5 shell electrons should also be partly responsible for the increase in eigenstate energies of inner shell electrons of implanted tin ion, because the outer electrons screen to some extent the nuclear charge seen by the inner shell electrons. The eigenstate energies of a neutral atom and the corresponding singly ionized ion have been tabulated in NIST Atomic Reference data for electronic structure calculations [24] for all elements of the Periodic Table. Using those tabulated numbers [24], the increases of eigenstate energies (1s,2s,3s,4s) of implanted tin ion in different lattice environments due to the screening effect of n=5 shell electrons have been estimated assuming that the screening effect of all n =5 electrons (s,p,d) are approximately equal and so somewhat overestimating the effect. After subtracting out this screening effect from the total predicted change of eigenstate energies by TB-LMTO code, we find that the remaining increases of 1s, 2s, 3s, 4s orbital eigenstate energies of tin implanted in gold versus lead are 23.6 eV, 23.8 eV, 22.6 eV and 18.1 eV respectively. We think that these remaining increases of eigenstate energies should be due to the different compressions experienced by the implanted ion in different lattice environments. Similarly the corresponding numbers for other cases have also been determined. Since an indium atom is bigger than a tin atom, so much larger changes of eigenstate energies were found for implanted indium in gold versus lead lattice. However TB-LMTO also predicts an increase of about 2.8 electrons in the n=5 valence shell of indium ion implanted in gold versus lead. Most of the additional valence electrons go to p-orbital and do not contribute to the electron density at the nucleus, but would increase the eigenstate energies of 1s, 2s, 3s, 4s orbitals of the indium ion. So in this case, the correction due to the screening becomes about 40% of the total increase of eigenstate energies and finally the remaining increases of eigenstate state energies of 1s, 2s,3s and 4s orbitals of indium due to the compression in gold versus lead lattice are 32.6 eV, 30.0 eV, 26.6 eV and 20.2 eV respectively.

Although the TB-LMTO code calculates the change of eigenstate energies of core electrons in different environments, it keeps the wave function of core electrons unchanged in frozen core approximation. It calculates the eigenstate energies of core electronic states by estimating the expectation values of the relevant Hamiltonian (that includes Hartree potential and electron-nucleus potential) using free atom core electronic wave functions. This approximation should be alright for the calculation of compressional energy, because the compression is primarily arising due to the repulsion among valence electronic orbitals and TB-LMTO code calculates wave functions of valence electrons rigorously. So the estimate of the compressional energy of different s-orbitals of the implanted ion can be used to determine the increase of the electron density at the nucleus of the implanted ion. However we cannot estimate the change of electron density at the nucleus due to compression using TB-LMTO code, because the code does not calculate the change of wave function under spatial confinement and so some other method has to be used. We are not aware of any standard code that considers both the change of electron density of core electrons and condensed matter effects. So we have made several simplifying assumptions to perform these calculations.

In order to estimate the changes of 1s, 2s, 3s, 4s electronic densities at the nucleus due to the increase of eigenstate energies as a result of the compressional effect on the implanted ion, we have used a simple Thomas-Fermi model of atom [25]. According to this model, the electrons are in an impenetrable spherical box (or bag) and they have only kinetic energy. The electron density is considered uniform throughout the bag. This picture is certainly a gross oversimplification, but still appears to be doing reasonably well for calculating the change of electron density at the atomic nucleus. The justifications for the use of Thomas-Fermi model of atom for performing this part of the calculation are based on the following assumptions.

1) The calculated compressional energies of different s-orbital eigenstates of the implanted ion are very small compared to the corresponding total energies of the s-orbital eigenstates. Since according to TB-LMTO formalism, the repulsion among valence electrons of neighboring atoms results in the hard sphere approximation and Hartree potential which is spherically symmetric around the atomic nucleus and arises due to the compression of the atom, so the compressional effect on inner shell electronic s-orbital of the atom should be isotropic. Hence for small change of volume, it is assumed that in the zeroth order, the functional form of s-orbital wave function remains almost unchanged due to the relatively small compression applied to the implanted ion and only the normalization constant of the wave function changes. As a result, the percentage change of the electron density at the nucleus is the same as the percentage reduction of the average volume of s-orbital in the lattice environment. So the detailed radial functional form of s-orbital wave function becomes unimportant for this calculation.

2) It is assumed that the compressional energy is primarily increasing the kinetic energy of the s-orbital electrons. In other word, the reduction of the average volume of s-orbital in the lattice is only causing an increase of the kinetic energy of s-orbital electrons.

3) It is assumed that in the zeroth order, the relationship between the increase of the kinetic energy of s-orbital electrons due to the corresponding decrease of the average volume of s-orbital is independent of the presence of a nuclear charge and its interaction with the electrons.

Actually, the increase of the kinetic energy of s-orbital electrons due to compression should modify the functional form of s-orbital wave function and it should tend to look more like particles in a box type wave function. Recently Aquino et al. [26] theoretically studied compressed helium atoms by placing a helium atom in an impenetrable spherical box and found that the ground state energy of the atom increases as the radius of the box is reduced. They also found [26] when the radius of the box is made smaller than 0.58 ×10$^{-8}$ cm i.e. about the size of Bohr radius of the atom, then it becomes more convenient to use free electrons in a box type basis states to describe the wave function of the helium atom. On the basis of the above-mentioned assumptions and Aquino et al.'s calculations [26], we think Thomas-Fermi model of atom can be used to calculate the increase of electron density at the nucleus due to the compression of the implanted ion in the zeroth order. However the effect of different s-orbitals for calculating electron densities at the nucleus has to be considered separately and proper weighting factors have to be applied. Since for both tin and indium, only 1s, 2s, 3s, 4s core electrons can have nonzero electron density at the nucleus and there should be different weighting factors for different orbitals for calculating electron density at the nucleus, so we consider four different electron bags each containing two electrons for 1s, 2s, 3s and 4s orbitals. Then using Thomas-Fermi model [25], the total energy of the electrons in a bag should be

$$E = \left(6.92 \times 10^{11}\right) \times \left(\frac{1}{V}\right)^{\frac{2}{3}} eV$$ at T=0 K, where V is the volume of the bag in cubic Fermi unit. We assume that the increase in eigenstate energy is associated with the reduction of the volume of the corresponding bag. So there would be corresponding increase of electron density throughout the bag. Differentiating the energy expression, we obtain

$$\left(\frac{dV}{V}\right) = -\left(0.217 \times 10^{-11}\right) \times (V)^{\frac{2}{3}} \times (\Delta E) \ldots\ldots\ldots\ldots(1)$$

The values of $<V^{2/3}>$ for the different bags are obtained from the corresponding uncompressed eigenstate (1s, 2s, 3s, 4s) wave functions of indium and tin as calculated by TB-LMTO code. The values of ΔE for different eigenstates are also obtained from TB-LMTO code and NIST calculations [24] as discussed earlier. In this way, the values of ΔV/V for 1s, 2s, 3s and 4s orbitals were obtained using eq.(1). Since the values of ΔE(1s), ΔE(2s), ΔE(3s), ΔE(4s) are similar, but the energy of higher orbitals decreases substantially, so as a result of compression, the percentage change of electron density of higher orbitals should be substantially higher. However because of the electronic structure of the atom, the contribution to electron density at the nucleus decreases rapidly for higher orbitals. These weighting factors for different orbitals were determined from the ratios of $\frac{|\psi_{2s}(r=0)|^2}{|\psi_{1s}(r=0)|^2}$, $\frac{|\psi_{3s}(r=0)|^2}{|\psi_{1s}(r=0)|^2}$ and $\frac{|\psi_{4s}(r=0)|^2}{|\psi_{1s}(r=0)|^2}$, where $\psi_{1s}(r=0)$, $\psi_{2s}(r=0)$, $\psi_{3s}(r=0)$, $\psi_{4s}(r=0)$ denote 1s, 2s, 3s, 4s orbital wave functions at the nucleus i.e. at r=0. These weighting factors for indium and tin were calculated using corresponding uncompressed wave functions as obtained from TB-LMTO code. Neglecting the contribution from 5s state, the change of electron density at the nucleus would be

$$\left.\frac{\Delta V}{V}\right|_{total} = \left.\frac{\Delta V}{V}\right|_{1s} + \left(\frac{|\psi_{2s}(r=0)|^2}{|\psi_{1s}(r=0)|^2}\right) \times \left.\frac{\Delta V}{V}\right|_{2s} + \left(\frac{|\psi_{3s}(r=0)|^2}{|\psi_{1s}(r=0)|^2}\right) \times \left.\frac{\Delta V}{V}\right|_{3s} + \left(\frac{|\psi_{4s}(r=0)|^2}{|\psi_{1s}(r=0)|^2}\right) \times \left.\frac{\Delta V}{V}\right|_{4s}$$

The increase in decay rate would be proportional to $\left.\frac{\Delta V}{V}\right|_{total}$. Using this procedure, we find the increase in decay rate of indium implanted in Au versus Pb is = 0.86% and for tin the corresponding number is = 0.67%. We also find that the increase in decay rate of indium implanted in Au versus Al is = 0.16% whereas for tin, the corresponding number is = 0.02%. These numbers are in reasonable agreement with our experimental results.

In our earlier works [4, 9, 10], we understood the change of decay rate of $^7$Be implanted in different media in terms of the change of valence 2s electrons of $^7$Be in different media and the corresponding change of electron density at the nucleus. We used the same method described earlier to determine the effect of compression on the decay rate of $^7$Be implanted in different lattice environments such as in Au versus $Al_2O_3$ lattices. According to TB-LMTO calculations, the eigenstate energy of 1s orbital of beryllium should increase by about 3.55 eV when implanted in Au versus $Al_2O_3$. TB-LMTO code also predicts that the number of n=2 valence electrons (including 2s, 2p etc.) of beryllium in Au (octahedral position) should be more than that of beryllium in $Al_2O_3$ by about 0.3 electrons, on the average. Using the tabulated results from NIST calculations [23], we find that the increase of 1s eigenstate energy of beryllium due to the screening by additional 0.3 number valence electrons is also about 3.5 eV. So the screening effect by the additional valence electrons can essentially account for the increase of 1s eigenstate energies of $^7$Be, as predicted by TB-LMTO code. Hence the difference in compressional effect on beryllium atom in Au and $Al_2O_3$ lattices is negligible. This result is very reasonable, because the small $^7$Be atom is not expected to experience any significant compression in normal lattice environment. So the change of decay rate of $^7$Be should be governed by the available number of valence 2s electrons of beryllium as discussed in earlier works [4,10].

Hensley et al.[14] observed that the decay rate of $^7$Be increases by 0.59% when 270 kilobar of pressure is applied on $^7$BeO causing about 10% reduction of the lattice volume. We performed TB-LMTO calculations by adjusting lattice parameters of BeO so that the lattice volume decreases by 10%. Then using TB-LMTO code, we find that 1s eigenstate energy of beryllium increases by 1.17 eV as a result of 10% compression of lattice volume. According to TB-LMTO calculation, the total number of n=2 electrons of beryllium remains about the same, when the lattice volume is reduced by about 10%. So the predicted 1.17 eV increase of 1s eigenstate energy of beryllium should be primarily due to the compression experienced by the beryllium ion as a result of the reduction of its muffin-tin radius. Using our method of treating 1s electrons in a bag containing two electrons in the framework of Thomas-Fermi model [25], we find that due to 1.17 eV increase of 1s eigenstate energy, the volume of the bag should decrease by about 0.89% and so the decay rate of $^7$Be should increase by about 0.89%, in reasonable agreement with the experimental result of 0.59%. It may be noted that although the lattice volume of BeO is reduced by about 10%, but the percentage increase of the kinetic energy of 1s orbital is very small and so the corresponding reduction of the volume of 1s orbital is only 0.89%. The percentage volume reduction of 2s orbital is much larger than that of 1s orbital, but it is much further away from the atomic nucleus and so its weighting factor is only 3.3% that of 1s orbital. So the increase of electron density at the nucleus due to the compression of valence 2s orbital is negligible compared to the effect of 1s orbital. In Table II, we show our comparison of experimental and theoretical results. In conclusion, we find that the decay rate of large radioactive atoms such as $^{109}$In and $^{110}$Sn can increase by observable amount due to the compression experienced by the atom in the lattice environment as a result of the spatial confinement. Earlier observations regarding the change of decay rate of $^7$Be can also be understood in our model.

**V. Summary**

We have measured the change in half-life of electron-capturing $^{109}$In and $^{110}$Sn nuclei implanted in different media and found that the decay rates of $^{109}$In and $^{110}$Sn increase by (1.00±0.17)% and (0.48±0.25)% respectively when implanted in the small Au lattice versus large Pb lattice. These observations cannot be understood in terms of the change in the number of valence electrons of indium and tin atoms implanted in different media, because the overlap of valence electrons of indium and tin at the nucleus (r=0) is negligible. However because of their large sizes, both indium and tin atoms should experience significantly higher compression in the small Au lattice compared to that in the large Pb lattice. As a result, the electron density at the nucleus and corresponding electron capture rate would increase when the atom is in the smaller Au lattice. We have understood our experimental results using TB-LMTO and Thomas- Fermi model calculations. The justifications for using simple Thomas-Fermi model of atom have been discussed. However our theoretical calculations should be considered a zeroth order calculation because of the many assumptions and approximations used in the calculations. In the case of implantation of small $^7$Be atoms in different media, the effect of lattice compression is generally negligible on such a small atom and so the change of decay rate of $^7$Be in different media is primarily due to the change of valence 2s electrons of beryllium in different media, as discussed in our earlier works [4,10].

## VI. Acknowledgements

We thank R. Vandenbosch (University of Washington, USA), K. Snover (University of Washington, USA), D. Storm (University of Washington, USA), S. Kailas (Bhabha Atomic Research Center, Mumbai, India) and A. Mookerjee (S. N. Bose Center for Basic Research, Kolkata, India) for useful discussions. We also thank the staff members of our cyclotron operation and data acquisition section for their support.

**Table-1**

Half-lives of $^{110}$Sn and $^{109}$In in different media

| Nucleus | Medium | Reduced $\chi^2$ of exponential fit | Half-life in hours |
|---------|--------|-------------------------------------|--------------------|
| $^{110}$Sn | Au | 2.11 | 4.1454±0.0062 |
|  | Pb | 2.08 | 4.1653±0.0085 |
|  | Al | 1.88 | 4.1559±0.0070 |
| $^{109}$In | Au | 1.98 | 4.1427±0.0042 |
|  | Pb | 1.67 | 4.1844±0.0057 |
|  | Al | 2.07 | 4.1514±0.0073 |

**Table 2**

Comparison between Experimental and Calculated values

| Decay rate difference between | Experimental (%) | Calculated (%) |
|-------------------------------|------------------|----------------|
| In in Au and In in Pb | 1.00±0.17 | 0.86 |
| In in Au and In in Al | 0.21±0.20 | 0.16 |
| In in Al and In in Pb | 0.80±0.22 | 0.58 |
| Sn in Au and Sn in Pb | 0.48±0.25 | 0.67 |
| Sn in Au and Sn in Al | 0.25±0.23 | 0.02 |
| Sn in Al and Sn in Pb | 0.23±0.25 | 0.52 |
| BeO (un compressed ) and BeO (270 kbar pressure) | 0.59±0.06 | 0.89 |

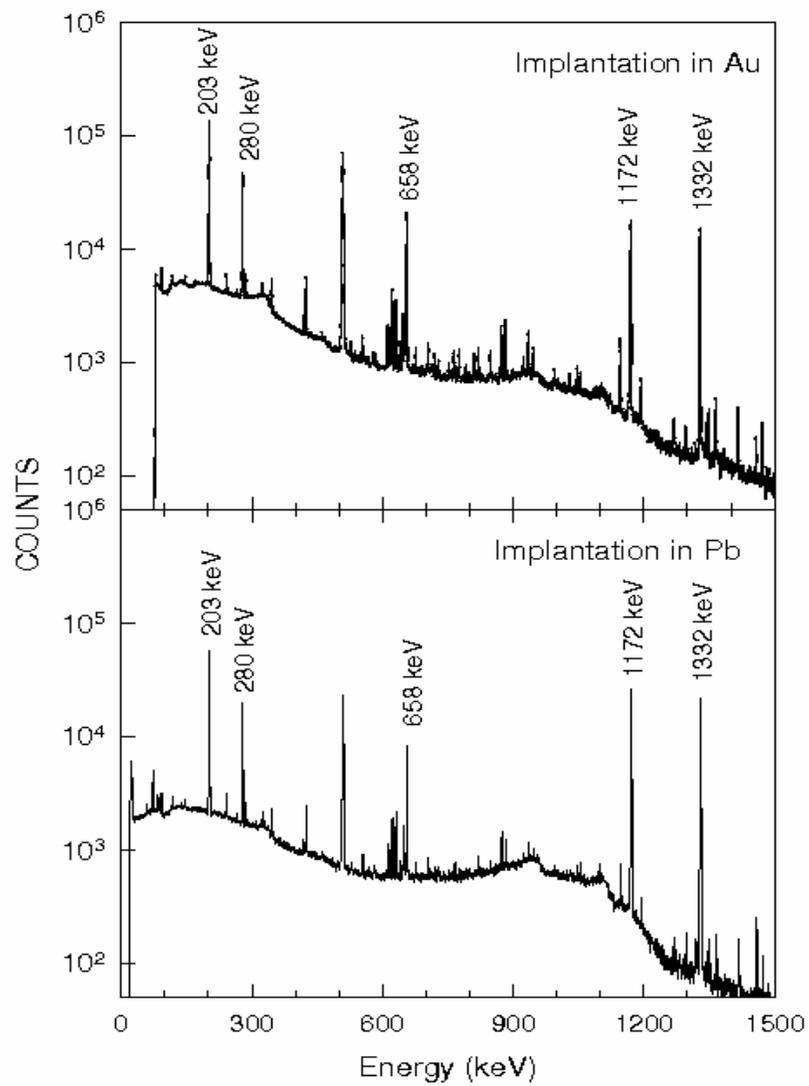

Fig. 1  Typical γ-ray spectra of sources in Au and Pb.

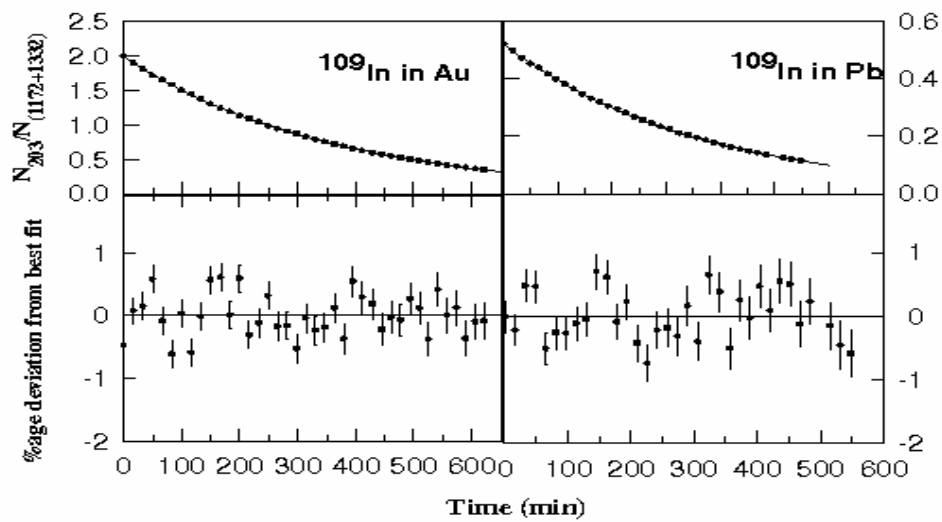

Fig 2. Exponential fits (upper panel) and residual plots (lower panel) of the data points for $^{109}$In in Au and Pb media.

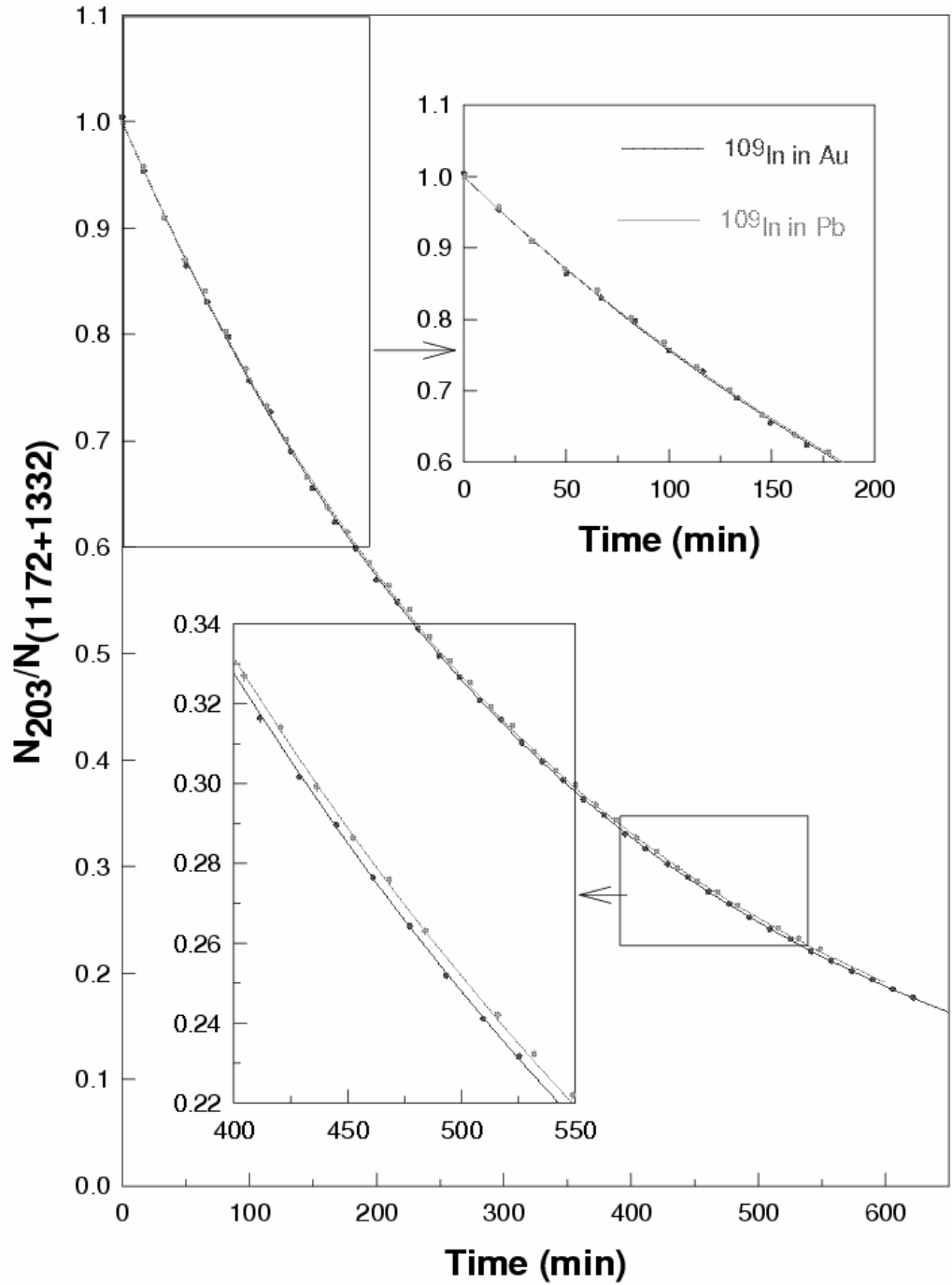

Fig. 3 Superimposition of exponential fits of $^{109}$In in Au and Pb. Time(minutes) has been plotted along X-axis. The ratio of peak area of the 203 keV γ-ray to the sum of the peak areas of 1173 keV and 1332.5 keV γ-rays of $^{60}$Co has been plotted along Y-axis. Error bars are smaller than the sizes of the data points. The method of normalizing the decay curves has been explained in the text.